\documentclass[useAMS,usenatbib]{mnras}

\usepackage{graphicx}
\usepackage{natbib}
\usepackage{amssymb}
\usepackage{amsmath}
\usepackage{setspace}
\usepackage{epsfig,color}
\usepackage{multicol}
\usepackage{hyperref}
\usepackage{upgreek} 
\usepackage{longtable}
\usepackage[normalem]{ulem}

\long\def\symbolfootnote[#1]#2{\begingroup
\def\thefootnote{\fnsymbol{footnote}}\footnote[#1]{##https://www.overleaf.com/project/5b98c85301ba633150bd62a42}\endgroup} 

\title{Density profile of a self-gravitating polytropic turbulent fluid in the context of ensembles of molecular clouds}
\title[Density profile of a turbulent fluid]{Density profile of a self-gravitating polytropic turbulent fluid in the context of ensembles of molecular clouds}

\author[Donkov et al.]
{
\parbox{\textwidth}{S. Donkov$^{1}$\thanks{E-mail: savadd@tu-sofia.bg}, I. Zh. Stefanov$^2$, T. V. Veltchev$^{3, 4}$ and R. S. Klessen$^{4, 5}$}\vspace{0.4cm} \\
\parbox{\textwidth}{
  $^1$Institute of Astronomy and National Astronomical Observatory, Bulgarian Academy of Sciences, 72 Tsarigradsko Shose,1784 Sofia, Bulgaria\\
  $^2$Department of Applied Physics, Technical University, 8 Kliment Ohridski Blvd., 1000 Sofia, Bulgaria \\
  $^3$University of Sofia, Faculty of Physics, 5 James Bourchier Blvd., 1164 Sofia, Bulgaria \\
  $^4$Universit\"at Heidelberg, Zentrum f\"ur Astronomie, Institut f\"ur Theoretische Astrophysik, Albert-Ueberle-Str. 2, 69120 Heidelberg, Germany \\
  $^5$Universit\"{a}t Heidelberg, Interdisziplin\"{a}res Zentrum f\"{u}r Wissenschaftliches Rechnen, Im Neuenheimer Feld 205, 69120 Heidelberg, Germany }
}

\date{Submitted 2021 January 18}

\pagerange{\pageref{firstpage}--\pageref{lastpage}} \pubyear{2021}

\begin{document}
\label{firstpage}
\maketitle

\begin{abstract}
We obtain an equation for the density profile in a self-gravitating polytropic spherically symmetric turbulent fluid with an equation of state $p_{\rm gas}\propto \rho^\Gamma$. This is done in the framework of ensembles of molecular clouds represented by single abstract objects as introduced by Donkov et al. (2017). The adopted physical picture is appropriate to describe the conditions near to the cloud core where the equation of state changes from isothermal (in the outer cloud layers) with $\Gamma=1$ to one of `hard polytrope' with exponent $\Gamma>1$. On the assumption of steady state, as the accreting matter passes through all spatial scales, we show that the total energy per unit mass is an invariant with respect to the fluid flow. The obtained equation reproduces the Bernoulli equation for the proposed model and describes the balance of the kinetic, thermal and gravitational energy of a fluid element. We propose as well a method to obtain approximate solutions in a power-law form which results in four solutions corresponding to different density profiles, polytropic exponents and energy balance equations for a fluid element. One of them, a density profile with slope $-3$ and polytropic exponent $\Gamma=4/3$, matches with observations and numerical works and, in particular, leads to a second power-law tail of the density distribution function in dense, self-gravitating cloud regions.
\end{abstract}

\section{Introduction}   \label{Sec_Introduction}

Molecular clouds (MCs) are the sites of star formation in galaxies. They are characterized by very low temperatures ($T \sim 10-30$~K) and consist mostly of molecular hydrogen well mixed with small amounts of dust \citep[see][for a review]{BP_ea_2020}. The complex physics of MCs is governed by gravity, supersonic turbulence, magnetic fields and -- in the general case -- an isothermal equation of state (EOS). Accretion from the surrounding medium and feedback from new-born stars and supernovae play an essential role in cloud's evolution \citep{MacLow_Klessen_2004, McKee_Ostriker_2007, KG_2016}. MCs display fractal structure in a large range of spatial scales 0.001 pc $\lesssim L \lesssim 100$~pc \citep{Elmegreen_1997, HF_2012} wherein mean density varies from about $10^2$ cm$^{-3}$ at $L \sim$ 100 pc up to $>10^5$ cm$^{-3}$ at the scales of pre-stellar cores ($L \leq$ 0.1 pc). In denser substructures (such as proto-stellar cores), the effective EOS of the gas $p_{\rm gas}\propto\rho^{\Gamma}$ changes from roughly isothermal ($\Gamma=1$) in the outer regions to that of a `hard polytrope' with $\Gamma>1$ \citep{Fed_Ban_2015,KNW_2011}. Rotation of proto-stellar objects and strong magnetic fields may also influence the observable characteristics of the material at these small scales.

\begin{figure} 
\begin{center}
\includegraphics[width=84mm]{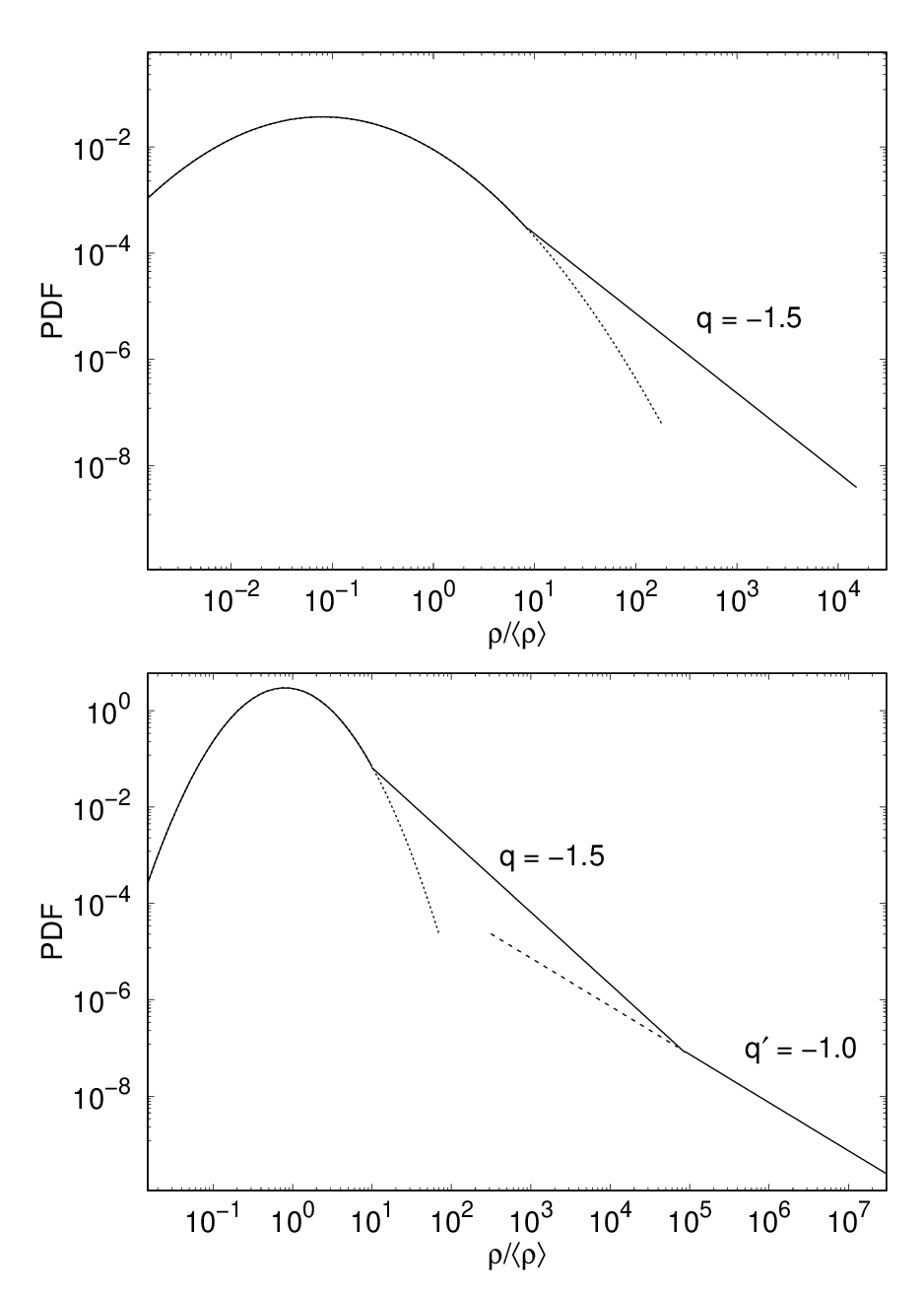}
\vspace{0.1cm}  
\caption{Examples of idealized evolved density PDFs with one (top) and two (bottom) PLTs. See text for the choice of slope values. The main part of the PDF is fitted with log-normal function (dotted).}
\label{fig_PDF_PLTs}
\end{center}
\end{figure}

One of the most important features of a MC is its probability density function (PDF). In this statistical characteristic are encoded the general structure and the evolutionary stage of the cloud. From observations one can obtain the PDF of column density $N$ ($N$-PDF) while numerical simulations of MCs enable derivation and analysis of the PDF $P(\rho)$ of mass density ($\rho$-PDF). To compare the two characteristics, one usually needs to assume that the observed cloud is nearly spherically symmetric and possesses a radial density profile  $\rho(l)\propto l^{-p}$ where $l$ is the given radius. At scales larger than 1 pc supersonic isothermal turbulence dominates the cloud physics and the $\rho$-PDF (hereafter, simply PDF) can be fitted well by a log-normal function, i.e., a Gaussian of log-density. This is well established for observational \citep{Kainu_ea_2009, Kainu_ea_2013, Lombardi_ea_2014, Schneider_ea_2015a, Schneider_ea_2015b, Schneider_ea_2016} and numerical $N$-PDFs \citep{Passot_VS_1998, Fed_ea_2008, Fed_ea_2010, Konstandin_ea_2012, Girichidis_ea_2014} as well for PDFs from simulations \citep{VS_1994,Passot_VS_1998,Kritsuk_ea_2007,Fed_ea_2010} and from theoretical considerations \citep{VS_1994, Passot_VS_1998, Kritsuk_ea_2007, Fed_ea_2010}. At smaller scales (and higher densities) above the scales of proto-stellar cores, the high-density part of the PDF gradually evolves from a log-normal wing to a power-law tail (PLT). In this density regime $P(\rho)\propto \rho^{q}$, with typical slope $-3 \ge q -1.5$ (Fig. \ref{fig_PDF_PLTs}, top) which corresponds to a radial density profile with $p=-3/q$, $1 \le p \le 2$ \citep[see][and the references therein]{DVK_2017}. 

The emergence of a PLT is usually explained with domination of gravity over turbulence at the corresponding small spatial scales \citep{Klessen_2000, Dib_Burkert_2005, Slyz_ea_2005, VS_ea_2008, KNW_2011, Collins_ea_2012}. Some examples of the PLT evolution from simulations and observations are discussed in \cite{Veltchev_ea_2019}. In general, the PDF slope $q$ gets shallower in the course of cloud evolution due to the formation of denser substructures. The apparent upper limiting value $q\simeq-1.5$ could be explained in the context of collapse of the so called singular isothermal spheres \citep{Penston_1969a, Penston_1969b, Larson_1969, Shu_1977, Hunter_1977, Whitworth_Summers_1985}, pressure-less gravitational collapse in strongly self-gravitating systems \citep{Girichidis_ea_2014},  scale-free gravitational collapse \citep{Li_2018}, collapse and dynamics of isolated gravo-turbulent cloud \citep{Jaupart_Chabrier_2020} or dynamical equilibrium between gravity and accretion \citep[][hereafter, Paper I and Paper II]{DS_2018, DS_2019}. Some numerical \citep{KNW_2011} and observational studies \citep{Schneider_ea_2015c} indicate existence of a {\it second} PLT corresponding to substructures at the scale of proto-stellar cores (Fig. \ref{fig_PDF_PLTs}, bottom). This second PLT is typically shallower, with a slope $q^\prime\simeq-1$ (corresponding to a density-profile exponent $p=3$) and hints at a slower accretion of gas from larger structures to the small-scale substructures in the cloud. There is still no clear physical explanation of this phenomenon. Some possible reasons might be rotation of the contracting core, strong magnetic fields or change in the thermodynamic EOS \citep{KNW_2011, Schneider_ea_2015c}.

In this paper we present a theoretical model of the density profile in the vicinity of the core of a self-gravitating spherically symmetric turbulent cloud. We suggest that the transition between the first and the second PLT of the cloud's PDF marks a change in the thermodynamic state of the gas from isothermal, at larger scales, to polytropic with exponent $\Gamma>1$ at scales comparable to the size of proto-stellar cores. On this assumption and starting from the equations of the gas medium under the condition of steady state in regard to macroscopic and microscopic motions, we obtain an equation for the conservation of energy of a fluid element per unit mass which is actually the Bernoulli equation for our model. Solving this equation up to the leading order terms we obtain four cases corresponding to different density profiles, polytropic exponents and energy balance equations for the fluid element. One of the solutions yields $p=3$ $(q=-1)$ and $\Gamma=4/3$ and we consider it as a possible explanation of the second PLT.

The paper has the following structure. In Section \ref{Sec-Set up of the model} we recall the model from Paper I and Paper II and introduce a change  in it regarding the thermodynamics near to the core of the cloud. Section \ref{Sec-Equ_rho(l)} is dedicated to the derivation of the equation for the density profile  in the following steps: comments on the equations of the medium (Section \ref{Subsec-equ of the medium}); derivation of the equation of energy conservation (per unit mass) of a moving fluid element (Section \ref{Subsec-equ_rho(l)_general}); comments on the explicit form of the terms in the latter equation  (Section \ref{Subsec-equ_rho(l)_terms}); writing down the explicit form of the equation for the density profile near to the cloud core (Section \ref{Subsec-equ_rho(l)_diff equ}). The derived equation is studied in Section \ref{Sec-analysis and sol} and its solutions in four different physical sub-cases are obtained. Section \ref{Sec-discussion} contains a discussion on the obtained solutions and on the model in general. Our conclusions are presented in Section \ref{Sec-conclusions}.

\section{Setup of the model}   
\label{Sec-Set up of the model}

%\begin{figure} 
%\begin{center}
%\includegraphics[width=70mm]{MC_class_of_equivalence.pdf}
%\vspace{0.1cm}  
%\caption{Concept of the MC class of equivalence (after \citealt{DVK_17}).}
%\label{fig_MC_class_of_equivalence}
%\end{center}
%\end{figure}

The cloud model was originally introduced in Paper I and developed further in Paper II. Here we review its basic features and modify its thermodynamics part.

The cloud is modelled as gaseous, spherically symmetric and self-gravitating ball. The matter accretes through its outer boundary of radius $l_{\rm c}$ and passes through all spatial scales down to a very small and dense core. The core is homogeneous with radius $l_0\ll l_{\rm c}$, density $\rho_0$ and mass $M_0$ which increases slowly in respect to the accretion time-scale.  

We assume a profile $\rho(\ell)$ of mass density which is related to the volume-weighted PDF $P(s)$ through the equation:

\begin{equation}
\label{rho(L)-p(s)}
P(s)ds=-3 \ell^2 d\ell~~~,~~~s\equiv\ln(\rho(\ell)/\rho_{\rm c})\equiv\ln(\varrho(\ell))~,
\end{equation}
where $\rho_{\rm c}=\rho (l_{\rm c})$ is the density at the outer cloud boundary and $\varrho\equiv\rho/\rho_{\rm c}$ is the dimensionless density. The dimensionless radius (scale) $\ell$ is related to the PDF by use of the integral form of Eq. (\ref{rho(L)-p(s)}):

\begin{equation}
\label{def_scale}
\ell(s)=\Bigg(\int\limits_{s}^{s_{0}}P(s')ds'\Bigg)^{1/3}~.
\end{equation}

The dimensionless characteristics of the core are $\ell_0=l_0/l_{\rm c}\ll 1$ and $s_0=\ln(\varrho_0)=\ln(\rho_0/\rho_{\rm c})$. Hence the spatial scales outside the core span the range $\ell_0\leq\ell\leq 1$.

The cloud is turbulent. We assume that turbulence is locally homogeneous and isotropic in each shell with radius $\ell$ and volume $dV=4\pi \ell^2 d\ell$. The inertial range of the turbulent cascade encloses all scales $\ell_0\leq\ell\leq 1$ and thus the dissipation can be neglected. The entire cloud (at every scale) is at steady state concerning both the macro-states (motions of the fluid elements) and the micro-states (thermal motion of the molecules).

In Paper II we extended the model further to account for the gravity of matter outside the cloud. We assume that the external gravitational potential $\varphi^{\rm ext}$ does not depend on the position of the fluid element moving through the cloud scales. This is a simplified assumption which holds only if the distribution of matter outside the cloud obeys a radial symmetry. The latter is not the case in vicinities of real clouds but it is in the spirit of the whole model. Another assumption of ours is that $|\varphi^{\rm ext}|$ and the absolute value of the potential, measured at cloud's boundary, are of the same order \citep[see][]{BP_ea_2018}.

In this paper we modify the thermodynamics of the cloud. The gas is considered isothermal in the outer shells \citep{Ferriere_2001}, far from the central high density region, while it obeys the EOS of `hard polytrope' near to the core where the gravitational potential of the latter becomes dominant in the energy balance. 

We point out that the model should be viewed not only as a convenient simplification of the real clouds, but rather as a generalization. The modelled cloud is treated as an average representative of the entire ensemble of clouds with one and the same PDF, size, density at the outer boundary and temperature (see \citealt{DVK_2017} and Paper I for more details). Our goal is not to reproduce in detail the cloud's morphology and dynamics but to derive the main physical properties typical for an entire ensemble of clouds.

\section{Derivation of the equation for density profile}
\label{Sec-Equ_rho(l)}

\subsection{Equations of the medium}
\label{Subsec-equ of the medium}

Now we aim to obtain an equation which determines the density profile $\rho(\ell)$ near to the core, within the model set above. The starting point are the equations of the medium:
\begin{itemize}
\item[-] The system of compressible Euler equations

\begin{equation}
\label{equ_cont}
\frac{\partial\rho}{\partial t} + \nabla\cdotp(\rho\vec{u})=0~~,
\end{equation}

\begin{equation}
\label{equ_N-St}
\frac{\partial\vec{u}}{\partial t} + \vec{u}\cdotp\nabla\vec{u} = -\frac{1}{\rho }\nabla p_{\rm gas} - \nabla\varphi~~,
\end{equation} 
with the first one reflecting the mass conservation and the second one describing the motion of a fluid element. It can be seen that the external force which introduces kinetic energy in the medium and the dissipative terms have been neglected. This is justified due to the statistical equilibrium which characterizes the inertial range of scales.

\item[-] The polytropic EOS of the gas

\begin{equation}
\label{equ_id-gas}
p_{\rm gas}=p_0 (\rho/\rho_{\rm c})^\Gamma~,~~~\Gamma>1~~,
\end{equation}
It reflects the assumption for the medium near to the core. Its form is chosen so that in the isothermal case ($\Gamma=1$) $p_{\rm gas}=c_{\rm s}^2 \rho$, where $c_{\rm s}$ is the isothermal sound speed \citep{Fed_Ban_2015}. Hence, $p_0=c_{\rm s}^2 \rho_{{\rm c}}$. This assumption allows us to connect continuously the two EOS of the gas: for the outer shells (isothermal gas) and for the inner shells (`hard polytrope').

\item[-] The Poisson equation for the gravitational potential

\begin{equation}
\label{equ_Pois}
\Delta\varphi=4\pi G \rho~~,
\end{equation}
which determines the gravitational potential produced by the given density distribution.
  
\end{itemize}

Other factors like angular momentum of the core, magnetic fields and feedback from the newborn stars have been neglected. This is done deliberately, to alleviate the first step towards an equation for $\rho(\ell)$ near to the core.

\subsection{General form of the equation of energy conservation for a moving fluid element}
\label{Subsec-equ_rho(l)_general}

Since the modelled cloud is an abstract object (representative of ensemble of clouds), the equation for $\rho(\ell)$ must be written in a form which accounts for the assumed symmetries and physics. First, we derive the equation of motion in dimensionless form, following the procedure from Paper I but implementing a polytropic EOS (Eq. \ref{equ_id-gas}):
\begin{itemize}
\item[i)] Implementation of polytropic EOS: From $\nabla p_{\rm gas}/\rho=c_{\rm s}^2 (\rho_{\rm c}/\rho) \nabla (\rho/\rho_{\rm c})^\Gamma=c_{\rm s}^2 \varrho^{-1} \nabla \varrho^\Gamma=...=c_{\rm s}^2 (\Gamma/(\Gamma-1)) \nabla \varrho^{\Gamma-1}$ one obtains	
\[ \frac{\partial\vec{u}}{\partial t} + \vec{u}\cdotp\nabla\vec{u} = -c_{\rm s}^2\frac{\Gamma}{\Gamma-1}\nabla\varrho^{\Gamma-1} - \nabla\varphi~~.\]
	
\item[ii)] Multiplication of the above expression by an infinitesimal displacement $d\vec{r}=\vec{u}dt$ in the direction of the vector field $\vec{u}$:
\[ \vec{u}\cdotp\frac{\partial\vec{u}}{\partial t}dt + (\vec{u}\cdotp\nabla\vec{u})\cdotp\vec{u}dt = -c_{\rm s}^2 \frac{\Gamma}{\Gamma-1}(\nabla\varrho^{\Gamma-1})\cdotp d\vec{r} - (\nabla\varphi)\cdotp d\vec{r}~~,\]
arriving easily at 
\[  \vec{u}\cdotp\frac{\partial\vec{u}}{\partial t}dt + (\vec{u}\cdotp\nabla\vec{u})\cdotp\vec{u}dt = \frac{d}{dt}(u^2/2)dt = d(u^2/2)~~.\]
	
\item[iii)] Introduction of dimensionless variables $v^2\equiv u^2/c_{\rm s}^2$  and $\phi\equiv\varphi/c_{\rm s}^2$ and rewriting of the equation:
\[ d(v^2/2) = -\frac{\Gamma}{\Gamma-1}(\nabla\varrho^{\Gamma-1})\cdotp d\vec{r} - (\nabla \phi)\cdotp d\vec{r}~~.\]
	
\item[iv)] Further modification of the equation. With the substitutions $d(\varrho^{\Gamma-1}) = (\partial (\varrho^{\Gamma-1})/\partial t)dt + (\nabla\varrho^{\Gamma-1})\cdotp d\vec{r}$ and $d\phi = (\partial\phi/\partial t)dt + (\nabla\phi)\cdotp d\vec{r}$, we obtain finally:
\[ d(v^2/2) = -\frac{\Gamma}{\Gamma-1}[d(\varrho^{\Gamma-1}) - (\partial (\varrho^{\Gamma-1})/\partial t)dt] - [d\phi - (\partial\phi/\partial t)dt]~~.\]
\end{itemize}

After these general steps, we constrain the consideration to the abstract representative of the cloud ensemble. The quantities which pertain to it are obtained after ensemble averaging with respect to the chaotic motion of the fluid elements in each shell and are put in brackets. By their use, the equation obtained at step v) above can be rewritten:
\begin{eqnarray}
\begin{aligned}
d\langle v^2/2 \rangle = -\frac{\Gamma}{\Gamma-1}[d\langle \varrho^{\Gamma-1} \rangle - (\partial \langle \varrho^{\Gamma-1} \rangle/\partial t)dt] \nonumber\\
- [d\langle \phi \rangle - (\partial \langle \phi \rangle/\partial t)dt]~.
\end{aligned}
\end{eqnarray}

At this point we take advantage of the assumption that the system is in steady state, i.e. $\partial \langle \varrho^{\Gamma-1} \rangle/\partial t = 0$ and $\partial \langle \phi \rangle/\partial t = 0$. The last equation takes the form:
\begin{equation}
\label{dE=0}
d\bigg[\langle v^2/2 \rangle + \frac{\Gamma}{\Gamma-1}\langle \varrho^{\Gamma-1} \rangle + \langle \phi \rangle \bigg]=0~.
\end{equation}

Since turbulence is, by assumption, locally homogeneous and isotropic it gives no contribution to the ensemble averaged motion of the fluid elements. The latter is indeed a radial in-fall towards the core which is at the centre of the ball. Due to the spherical symmetry there is a one to one correspondence between radial position and scale. As mentioned in point ii) above, the total differential represents a shift in the direction of motion of the fluid element and, hence, a shift along the scales $\ell$. This means that the obtained Eq. (\ref{dE=0}) is a differential form  of the law of conservation of energy per unit mass of a fluid element as it passes from scale $\ell$ to scale $\ell+d\ell$. Moreover, the abstract object is characterized by an averaged density profile -- the latter is related to the PDF, which is averaged over the cloud ensemble per definition. Then, $\langle\varrho^{\Gamma-1}\rangle=\varrho^{\Gamma-1}$. With these considerations we obtain an equivalent form of equation (\ref{dE=0}):
\begin{equation}
\label{dE/dl=0}
\frac{d}{d\ell}\bigg[\langle v^2/2 \rangle + \frac{\Gamma}{\Gamma-1}\varrho^{\Gamma-1} + \langle \phi \rangle \bigg]=0~.
\end{equation}

Although the turbulent velocity plays no role in the averaged motion of the fluid element, it has a non-vanishing contribution to the kinetic energy term $\langle v^2/2 \rangle$ due to the scalar nature of the latter.

\subsection{Explicit form of the terms in equation (\ref{dE/dl=0})}
\label{Subsec-equ_rho(l)_terms}

In this subsection we derive the explicit form of the terms in equation (\ref{dE/dl=0}), making use of some considerations and results from Papers I and II. As shown in Paper I (see Section 3.2 there), the kinetic energy term satisfies:
\begin{equation}
\label{v=vt+va}
\langle v^2 \rangle = \langle v_{\rm t}^2 \rangle + \langle v_{\rm a}^2 \rangle~,
\end{equation}
where $\langle v_{\rm t}^2 \rangle$ is the turbulent kinetic energy per unit mass and $\langle v_{\rm a}^2 \rangle$ is the accretion kinetic energy per unit mass. 

In general, turbulent velocity fluctuations in molecular clouds obey a power-law scaling relation $u=u_0 (L/1~\mathrm{pc})^{\beta}$ \citep{Larson_1981, Padoan_ea_2006, Kritsuk_ea_2007, Fed_ea_2010}, with some normalizing factor $u_0$ and scaling index $0\leq\beta\leq1$. In our model, the spherically symmetric cloud is ensemble averaged and we can apply such scaling relation for  $\langle v_{\rm t}^2 \rangle$:
\begin{eqnarray}
\label{vt2-p(s)}
\langle v_{\rm t}^2 \rangle = \frac{u_0^2}{c_{\rm s}^2}\bigg(\frac{l_{\rm c}}{pc}\bigg)^{2\beta} \ell^{2\beta} = T_0 \ell^{2\beta}~,
\end{eqnarray}
where $T_0\equiv (u_0^2/c_{\rm s}^2)(l_{\rm c}/{\rm pc})^{2\beta}$ is the ratio of the turbulent kinetic energy per unit mass of a fluid element at the cloud boundary to the thermal energy per unit mass, in case of the isothermal EOS assumed for the outer shells of the cloud.

No particular mechanism of interaction between accretion and turbulence is implied in this work (like in Paper II) -- rather, they are treated as formally independent. Our model is focused on integral characteristics of the cloud, such as the density PDF, and this justifies the approach of using scaling relations for the turbulent velocity with a power-law exponent in the range $0 \le \beta \le 1$. Various different ways of generating and driving turbulence have been discussed, e.g., by \citet{KH_2010}, \citet{Robertson_Goldreich_2012}, \citet{Xu_Lazarian_2020}, or \citet{GG_VS_2020}.

The explicit form of the accretion kinetic term $\langle v_{\rm a}^2 \rangle$ has been obtained from the continuity equation in Paper I (see Section 3.3 there). The main steps are as follows. The continuity equation (Eq. \ref{equ_cont}) is averaged with respect to the ensemble of the micro-states of the abstract object:
\[ \frac{\partial\rho}{\partial t} + \nabla\cdotp\langle\rho\vec{u}\rangle=0~.\]
The assumption of steady state, $\partial\rho/\partial t =0$, leads to:
\[ \nabla\cdotp\langle\rho\vec{u}\rangle=0~.\]
Taking advantage of the spherical symmetry of the vanishing turbulent velocity when the motion of a fluid element is averaged and also of the averaged (by assumption) density profile, we obtain: 
\[ \ell^4\varrho^2\langle v_{\rm a}^2 \rangle=A_0=\mathrm{const}(\ell)~,\]
and, hence, a formula for $\langle v_{\rm a}^2 \rangle$:
\begin{equation}
\label{va2-rho(l)}
\langle v_{\rm a}^2\rangle = A_0 \varrho(\ell)^{-2} \ell^{-4}~.
\end{equation}
The dimensionless coefficient $A_0$ is the ratio of the accretion kinetic energy term at the cloud boundary to the isothermal kinetic energy per unit mass.

The second term in Eq. (\ref{dE/dl=0}) represents the thermal potential and it relates to the thermal pressure, i.e. to the energy density related to the random thermal motion of gas particles. 

The third term in Eq. (\ref{dE/dl=0}) is the averaged gravitational potential of a fluid element at radius $\ell$. It is given by:
\begin{equation}
\label{grav-potencial}
\langle \phi \rangle= -\frac{G}{l_{\rm c} c_{\rm s}^2}\frac{M(\ell)}{\ell} -\frac{G}{l_{\rm c} c_{\rm s}^2}\frac{M_0}{\ell} + \langle \phi^{\rm ext} \rangle~,
\end{equation}
where $M(\ell)=3M_{\rm c}^{*}\int_{\ell_0}^{\ell} \ell'^2\varrho(\ell')d\ell'$ is the mass of the inner shells in respect to $\ell$ (excluding the core) and $M_{\rm c}^{*}= (4/3)\pi l_{\rm c}^3 \rho_{\rm c}$ is a normalizing coefficient (for its physical interpretation, see Section 3.2 in Paper I). The first term is the gravitational potential caused by the inner shells and the second one is the gravitational potential caused by the core at scale $\ell$. We note again that $M_0$ is assumed to increase slowly due to the accretion onto the cloud core and therefore its change in time is neglected. The last term in Eq. (\ref{grav-potencial}) reads: 
\[ \langle \phi^{\rm ext} \rangle = -(3G M_{\rm c}^{*}/l_{\rm c} c_{\rm s}^2)\int_{\ell}^1 \ell'\varrho(\ell')d\ell' + \varphi^{\rm ext}/c_{\rm s}^2~,\] 
where the first addend is the gravitational potential caused by the outer shells in respect to $\ell$ and the second one is the potential caused by the matter outside the cloud. We assume that all matter outside the cloud causes a potential term within the cloud $\varphi^{\rm ext}$ which does not depend on the position of the considered fluid element moving through the spatial scales. This constant term does not contribute to the element's dynamics and will be neglected hereafter. 

Eventually, the averaged dimensionless potential $\langle \phi \rangle$ can be expressed by the density profile $\varrho(\ell)$:

\begin{eqnarray} \label{grav-potencial_rho(l)}
\begin{aligned}
\langle \phi \rangle= -\frac{3G}{c_{\rm s}^2}\frac{M_{\rm c}^{*}}{l_{\rm c}} \frac{1}{\ell} \int\limits_{\ell_0}^{\ell} \ell'^2\varrho(\ell')d\ell' -\frac{3G}{c_{\rm s}^2}\frac{M_{\rm c}^{*}}{l_{\rm c}} \int\limits_{\ell}^1 \ell'\varrho(\ell')d\ell' \\
-\frac{G}{c_{\rm s}^2} \frac{M_0}{l_{\rm c}} \frac{1}{\ell}~.
\end{aligned}
\end{eqnarray}

\subsection{Derivation of the equation for $\varrho(\ell)$}
\label{Subsec-equ_rho(l)_diff equ}

Now Eq. (\ref{dE/dl=0}) can be written in a form which allows us to derive an equation for the density profile: 
\begin{eqnarray} \label{dE/dl=0_rho}
\begin{aligned}
\frac{d}{d\ell}\Bigg[A_0 \varrho(\ell)^{-2}\ell^{-4}+ T_0 \ell^{2\beta} + 2\frac{\Gamma}{\Gamma-1}\varrho^{\Gamma-1} \\
-\frac{3G_0}{\ell} \int\limits_{\ell_0}^{\ell} \ell'^2\varrho(\ell')d\ell' -3G_0\int\limits_{\ell}^1 \ell'\varrho(\ell')d\ell' -\frac{G_1}{\ell} \Bigg]=0~,
\end{aligned}
\end{eqnarray}
where the dimensionless coefficients $G_0=(2G/c_{\rm s}^2)(M_{\rm c}^{*}/l_{\rm c})$ and $G_1=(2G/c_{\rm s}^2)(M_0/l_{\rm c})$ are the ratio of the gravitational energy of the shells (excluding the core) to their thermal energy per unit mass and the ratio of the gravitational energy per unit mass of the core to its thermal energy per unit mass, respectively (see Paper I, Section 3.4). The expression in the parentheses in Eq. (\ref{dE/dl=0_rho}) is the total energy per unit mass of a fluid element. Denoting it by $E_0$, we get: 
\begin{eqnarray} \label{E_rho}
\begin{aligned}
A_0 \varrho(\ell)^{-2}\ell^{-4}+ T_0 \ell^{2\beta} + 2\frac{\Gamma}{\Gamma-1}\varrho^{\Gamma-1} -  \frac{3G_0}{\ell}\int\limits_{\ell_0}^{\ell} \ell'^2\varrho(\ell')d\ell' \\ 
-3G_0\int\limits_{\ell}^1 \ell'\varrho(\ell')d\ell' -\frac{G_1}{\ell}=E_0=\mathrm{const}~.
\end{aligned}
\end{eqnarray}
This is a non-linear integral equation for the dimensionless density profile $\varrho(\ell)$. 

\section{Study of the derived equation for density profile}
\label{Sec-analysis and sol}

A general approach is to find a solution of the equation in form of a Frobenius series \citep[see, e.g.][]{Riley_Hobson_Bence_2006} with increasing exponents and small $\ell$: $0<\ell_0<\ell\ll1$. In this work we search for it in the form of a power-law density profile $\varrho(\ell)=\ell^{-p}$ since the latter is in one-to-one correspondence to a power-law PDF $P(s)\propto \exp(qs)~,~s=\ln(\varrho)$ with $q=-3/p$ \citep{DVK_2017}. The motivation for this ansatz is to suggest a physical explanation of the second PLT observed in several nearby molecular clouds \citep{Schneider_ea_2021}. 

Making this substitution in Eq. (\ref{E_rho}), we arrive at:
\begin{eqnarray} \label{Ealg_rho}
\begin{aligned}
A_0 \ell^{2p-4} + T_0 \ell^{2\beta} + 2\frac{\Gamma}{\Gamma-1}\ell^{-p(\Gamma-1)}\\ 
- 3G_0 \frac{\ell^{2-p}}{3-p} \Bigg[1-\bigg(\frac{\ell_0}{\ell}\bigg)^{3-p}\Bigg] \\
- 3G_0 \frac{1-\ell^{2-p}}{2-p} - G_1 \ell^{-1} = E_0=\mathrm{const}~.
\end{aligned}
\end{eqnarray}

No value of the density-profile exponent $p$ yields an exact solution of this equation. Different assumptions and approximations yield different approximate solutions for the slope of the density profile $p$. 

Unlike Paper I and Paper II we study here only the case when the considered fluid element is near to the core, i.e. $\ell_0 \lesssim \ell \ll1$. One can immediately conclude that the gravitational term, accounting for the potential of the inner shells, is negligible due to the vanishing expression in the parentheses. This holds even for profiles as steep as $p=3$, because in this case one obtains a finite limit through L'Hospital's rule: 
\[ \frac{\ell^{2-p}}{3-p} \Bigg[1-\bigg(\frac{\ell_0}{\ell}\bigg)^{3-p}\Bigg]~~~\longrightarrow~~~~ -\ell^{-1}\ln(\ell_0/\ell)\ll \ell^{-1} ~.\]
Then Eq. (\ref{Ealg_rho}) can be simplified to:

\begin{eqnarray} \label{Ealg_rho_G1}
\begin{aligned}
A_0 \ell^{2p-4} + T_0 \ell^{2\beta} + 2\frac{\Gamma}{\Gamma-1}\ell^{-p(\Gamma-1)} - 3G_0 \frac{1-\ell^{2-p}}{2-p}\\ - G_1 \ell^{-1} = E_0~.
\end{aligned}
\end{eqnarray}

Furthermore, near to the core the gravity of the latter is important. Then the term with the smallest exponent $G_1\ell^{-1}$ must be of leading order (i.e., the dominant exponent is ``$-1$'') since $0<\ell<1$. To obtain a non-trivial solution, one needs to consider the balance between gravity of the core and, possibly, the influence of the outer shells on the motion of the fluid element (the negative terms involving $G_0$ and $G_1$) and the first three terms which are positive. The turbulent kinetic term is not of leading order because $2\beta\geq0$. Viable possibilities stem from the accretion kinetic term $A_0 \ell^{2p-4}$, the thermal term $(\Gamma/(\Gamma-1))\ell^{-p(\Gamma-1)}$ or their combination. Below we analyse those cases:

\begin{itemize}
	
	\item[$\bullet$] The accretion kinetic term is important when $2p-4=-1~~\Leftrightarrow~~p=3/2$. Then gravity of the outer shells is not important since $2-p=1/2>0>-1$. On the other hand, the thermal term can be important if and only if $-p(\Gamma-1)=-1~~\Leftrightarrow~~\Gamma=5/3$. This leads us to two sub-cases:\vspace{6pt}
	
	(1) $p=3/2$ and $\Gamma=5/3$, turning Eq. (17) into:
	\[ A_0 + 2\frac{\Gamma}{\Gamma-1} - G_1 = 0~~, \] 
	i.e. {\it both the accretion kinetic term and thermal term provide energy balance against the core gravity}. \vspace{6pt}
	
	(2) $p=3/2$ and $1<\Gamma<5/3$, leading to:
	\[ A_0 - G_1 = 0~~, \] 
	i.e. {\it only the accretion kinetic term provides energy balance against the core gravity}.\vspace{6pt}
	
	\item[$\bullet$] The thermal term is important and the accretion term negligible when $-p(\Gamma-1)=-1$ and $p>3/2$. In this case one gets for the polytropic index $1<\Gamma<5/3$ and $3/2 < p < \infty$ for the density-profile exponent. Two typical values for hard polytropes in this range, $\Gamma=3/2$ and $\Gamma=4/3$, yield the further sub-cases: \vspace{6pt}
	
	(3) $p=2$ and $\Gamma=3/2$ with
	\[ 2\frac{\Gamma}{\Gamma-1} - G_1 = 0~~, \] 
	i.e. {\it the thermal term provides the energy balance against the core gravity}\footnote{In this case we apply L'Hospital's rule for the gravitational term which accounts for the outer shells, with respect to the fluid element, and obtain: $$-\frac{1-\ell^{2-p}}{2-p} \longrightarrow \ln(\ell) \ll \ell^{-1}$$}.\vspace{6pt} 
	
	(4) $p=3$ and $\Gamma=4/3$ resulting in:
	\[ 2\frac{\Gamma}{\Gamma-1} - 3G_0 - G_1 = 0~~, \] 
	i.e. the thermal term provides the energy balance against the gravity, both of the core ($G_1$ term) and of the outer shells ($3G_0$ term).\vspace{6pt}
	
	It is worth to note here that Eq. (\ref{Ealg_rho_G1}) has no solutions in the range $\infty>p>3$ ($1<\Gamma<4/3$) -- in this case the term accounting for the gravity of the outer shells is of leading order (with exponent $2-p<-1$) and there is no term to balance it. This limits, in case the accretion term is not dominant, the possible ranges of the polytropic index and of the density-profile exponent to $4/3\leq \Gamma < 5/3$ and $3\geq p > 3/2$, respectively. Note that 
	$\Gamma=4/3$ is the critical exponent for the stability of stellar polytropes.
\end{itemize}

\section{Discussion}
\label{Sec-discussion}

\subsection{Towards a fiducial model}
\label{Subsec-discussion on the results}

The solutions obtained above exist only if the coefficients in Eq. (\ref{Ealg_rho_G1}) $A_0,~2\Gamma/(\Gamma-1),~3G_0$, and $G_1$ are of the same order. In the sub-cases (1), (3) and (4) one can assess this order using the coefficient $2\Gamma/(\Gamma-1)$ and taking a value of the polytropic index in the range $4/3 \leq \Gamma \leq 5/3$. This simple calculation gets us $8 \geq 2\Gamma/(\Gamma-1) \geq 5$, i.e. the coefficients are of order between $1$ and $10$. The sub-case (2) is an exception because the coefficient $2\Gamma/(\Gamma-1)$ does not play a role. However, $G_1=(2G/c_{\rm s}^2)(M_0/l_{\rm c})$ depends obviously on the model parameters and should be the same in all sub-cases. The same is true for $A_0$.

What conclusions regarding the modelled abstract cloud can be drawn out of the result that the coefficients $A_0$, $3G_0$ and $G_1$ are of order of a few? $G_0$ and $G_1$ are sophisticated functions of several model parameters: the temperature in the outer cloud shells (through $c_{\rm s}^2$), the cloud size $l_{\rm c}$, the density at the outer cloud edge (through $M_{\rm c}^{*}$) and the core mass $M_0$. However, $A_0$ has a simpler meaning -- it is, by definition, the ratio of the accretion kinetic energy term at the cloud boundary to the isothermal kinetic energy per unit mass. Hence the accretion velocity must be of the order of the sonic speed. This has implications for the applicability of our model to substructures in the cold neutral medium.

A further step in the analysis of the obtained solutions is to assess their consistency with a realistic power-law PDF. We recall here our objective to reproduce the density PDF of molecular gas in the a region close to the central proto-star/proto-cluster where the EOS switches from isothermal to one of `hard polytrope'. Some observations \citep{Schneider_ea_2015c} and simulations \citep{KNW_2011, Veltchev_ea_2019, Marinkova_ea_2021} indicate the emergence of a second PLT of the PDF, with slope $q\sim-1$. The latter corresponds to an exponent $p=3$ in the radial density profile which is combined with polytropic index $\Gamma = 4/3$ in the sub-case (4) obtained in the previous Section. Hence $2\Gamma/(\Gamma-1) = 8$ and from the equation of energy balance in this sub-case one gets:  
\[ 3G_0 + G_1 = 8~.\]

An independent assessment of these coefficients could be done from estimates of the physical characteristics of the cloud core. Per definition, $G_0=(2G/c_{\rm s}^2)(M_{\rm c}^{*}/l_{\rm c})$ and substituting $M_{\rm c}^{*} = (4/3)\pi \rho_{\rm c} l_{\rm c}^3$ (see Paper I) we have $G_0=(2G/c_{\rm s}^2)(4/3)\pi \rho_{\rm c} l_{\rm c}^2$. In the shells from the cloud edge down to the very close vicinity of the core the density profile is $\rho(l)=\rho_{\rm c}(l/l_{\rm c})^{-2}$. Then $\rho_{\rm c} l_{\rm c}^2 \approx \rho_0 l_0^2$ and hence:
\[ G_0\approx(2G/c_{\rm s}^2)(4/3)\pi \rho_0 l_0^2~. \] 
At this point we make use of the classical result of \citet[][see Fig. 1 there]{Larson_1969}. It provides estimates of density $\rho_0\sim 10^{-13} {\rm g}/{\rm cm}^3$ near to the opaque core of collapsing gas ball where the EOS switches from isothermal to one of `hard polytrope' with index $\Gamma\approx 4/3$ (see comments in Section 4 there), and of core size $l_0\sim 10^{14} {\rm cm}$. Then, adopting $T\sim 10~{\rm K}$ and $c_{\rm s}\approx 2\times10^4~{\rm cm}/{\rm s}$, finally we obtain $G_0\approx 1.4$. Thus both coefficients $3G_0\approx 4.2$ and $G_1\approx 8-4.2=3.8$ are of the same order and in the expected range between $1$ and $10$. Although there are differences between Larson's model and our model, these assessments hint that the first steps toward reproduction of the density PDF in regions close to cloud core are in the right direction.

\subsection{Model assumptions and obtained solutions: physical analysis}
\label{Subsec-physical analysis}

The presented model is built upon several main assumptions. Though closely related, they can be divided in two groups: geometrical and physical. The basic geometrical assumptions are the spherical symmetry of the cloud and the radial gas flow. This picture is far from the shape and dynamics of real clouds. It should be considered as a justified simplification in the attempt to derive the density profile of a cloud ensemble. A main characteristic of the latter is the volume-weighted PDF of a power-law type which leads -- in a natural way, -- to a one-dimensional model construction (see equation \ref{def_scale}) and a radial flow from the outer shells toward the inner, denser ones.

The main physical assumptions concern gravity, accretion, thermodynamics and turbulence in the cloud. Gravity and accretion are treated in consistence with the adopted geometry: a spherically symmetric gravitational field (both for self-gravity and for the material outside the cloud) and a radial steady-state accretion. The assumption for a steady-state flow is in a good agreement with numerical works \citep{KNW_2011, Veltchev_ea_2019, Marinkova_ea_2021} which indicate approximately constant slopes for the PDF tails at late stages of MC evolution. 

In regard to thermodynamics and turbulence, we distinguish between two physical regimes: shells located near to the core ($\ell_0\lesssim\ell$) and far from it ($\ell_0\ll\ell$). In the latter case, the equation of state is assumed to be isothermal ($\Gamma=1$) as justified from numerous observations and simulations \citep{Ferriere_2001}. In the regime near to the core we adopt an equation of state of a `hard polytrope' ($\Gamma>1$) -- in contrast to the setup in Papers I and II but in agreement to the expected conditions near to the centre of collapsing prestellar cores \citep{Larson_1969, Horedt_2013}. Turbulence is important in our model in view of the assumed steady-state flow and the neglected dissipation. It may contribute to the energy balance for the solutions far from the core (see Paper II, Section 3.1) but is not a leading-order term in the solutions near to it (Papers I \& II, this study). The role of turbulence is a matter of debate and there are two competing scenarios of cloud evolution which have been developed in the last two decades. In the gravo-turbulent (GT) scenario \citep{MacLow_Klessen_2004,HF_2012,KG_2016}, turbulence is a major physical factor acting across a wide range of scales, from giant MCs as a whole, where it provides support against gravitational collapse, down to very small scales within a cloud, where it is believed to trigger the formation of prestellar cores and foster star formation. This defines the inertial range\footnote{Where the dissipation is negligible.} of the turbulent cascade \citep[$\sim 100$ to $\sim 0.001$ pc;][]{Elmegreen_1997, HF_2012}; at its upper end the flows are strongly supersonic, with Mach numbers $\gtrsim 10$, while at the lower end they are moderately supersonic, transonic or even subsonic. Turbulence in the GT scenario is driven by external forces like accretion, supernova explosions, bipolar outflows, etc. In contrast, the global hierarchical collapse (GHC) scenario \citep[see][and the references there in]{VS_ea_2019} interprets the large-scale non-thermal motions oberved in the interstellar medium as being caused by gravity leading to hierarchical and chaotic collapse \citep{BP_ea_2011a, BP_ea_2011b}. According to the GHC scenario turbulence is only moderately supersonic (Mach numbers $\lesssim 3$) even at large cloud scales and does not provide support against gravity. It is inherited from the very process of cloud formation as converging flows in the warm neutral medium form cold large-scale structures through collisions. In this way, turbulence in the GHC scenario is a second-order factor in the cloud evolution -- at small scales it still causes the formation of clumps which collapse further to stars but this process is part of an energy and mass cascade driven by gravity.

Our model is consistent with both discussed scenarios in regard to some basic elements of its construction. On the one hand, like in the GT scenario, the spatial scales in consideration fall within the turbulent inertial range, i.e. the dissipation processes are neglected (see equation \ref{equ_N-St}). On the other hand, the mass accretion is postulated to take place across all scales -- which fits into the GHC scenario. As for the solutions found from the model, turbulence may play a role in the regime far from the core in case the velocity dispersion does not scale (i.e. $\beta=0$; see Paper II, Section 4.1) which refers to the so called ``coherent cores'', first mentioned in \cite{Goodman_ea_1998}. In the near-to-the-core regime turbulence is not a term of leading order which is expected in view of the small size of the protostar vicinity. In agreement with the GT as well with the GHC scenario, the accretion flow here is dominated by self-gravity.
  
All four solutions of equation (\ref{Ealg_rho_G1}) seem feasible in regard to the exponents of the density profile $1<p\leq3$ and of the polytropic index $1<\Gamma\leq5/3$. The density profile is shallower ($p=3/2$) in subcases (1) and (2) where the accretion term is important. Vice versa, in the subcases (3) and (4), which are characterized by steeper density profiles ($p>3/2$), accretion does not play a role in the energy balance; subcase (4) yields a slope $-1$ of the density PDF  corresponding to a second power-law tail at its high-density end found from high-resolution simulations of star-forming clouds \citep{KNW_2011, Marinkova_ea_2021}. So the most plausible solution in this set is the one with the steepest density profile, which provides the strongest support against accretion. It is consistent with the concept that the accretion slows down significantly near to the centre of a protostellar core. We can also speculate that the solutions (2), with $p=3/2$ and $1<\Gamma<4/3$\footnote{The range $4/3\le\Gamma<5/3$ is not relevant for this consideration.}, and (4), with $p=3$ and $\Gamma=4/3$, may be interpreted as descriptions of contiguous stages of the evolution of the second power-law tail. Note that the density profile in solution (2) reproduces the solution in the regime near to the core when adopting an isothermal EOS obtained in Paper II (see Section 4.2 there); the energy balance in both solutions is $A_0-G_1=0$ which corresponds to a free-fall collapse. One can imagine that as long as the gas is transparent for radiation and the EOS is isothermal, the energy balance yields a solution with $p=3/2$ -- a second power-law tail emerges in the density PDF. As material accumulates due to accretion, the medium becomes more denser and $\Gamma$ increases slowly from $1$ to $4/3$ while the density profile remains unchanged. When the gas becomes opaque enough, with $\Gamma\simeq4/3$, the density profile gradually steepens and reaches a slope $p=3$. This corresponds to a second power-law tail with slope $-1$ in the PDF which is found from the above-mentioned simulations and hence has to be indicative of a physically stable configuration. All this process must be very slow in regard to the characteristic timescale in the cloud in order to preserve the model assumption for steady state.

The solution (2) with $p=3/2$ and an energy balance equation for the fluid element $A_0-G_1=0$ (as well the identical one with isothermal EOS in Paper II) is similar to the theoretical result of \citet{Li_2018}. However, this author obtained $p=2$ for the density profile. This apparent contradiction can be explained with the different range of scales in consideration -- the solution of \citet{Li_2018} is found at the scales where the gravity of the central object is not important, whereas the solutions here and in Paper II are obtained in the regime near to the core (i.e. the central object in our model). Note also that \citet{Li_2018} does not consider any effects of thermodynamics to derive the density profile.

In contrast we suggest an explanation of the second power-law tail in the density PDF of MCs based on a change in thermodynamics. Alternative explanations could be, e.g., rotation of the core (protostar) or strong magnetic fields which can oppose the gravity and hence slow down the accretion \citep{KNW_2011, Schneider_ea_2015c}, due to centrifugal forces or magnetic tension. Most likely, all three factors act together, probably with varying magnitudes in specific cases. Our consideration is a simplified possible case which includes only a change in the EOS and shall be conceived as a first step in the exploration of a complex problem.

\subsection{Caveats}
\label{Subsec-debate on the model}

Two possible caveats to our model should come into discussion. The first one is the assumption of spherical symmetry of our cloud, which allows us to simplify the equations and the calculations. Real molecular clouds are far from being isotropic balls. Therefore we emphasize that the modelled object must not be viewed as a description of specific individual cloud but rather as an average representative of an ensemble of MCs with the same PDF, size, density at the outer boundary and temperatures. The considered object is also averaged in time and thus turbulence, which is locally homogeneous and isotropic by assumption, is not inconsistent with spherical symmetry. Moreover, our aim is not to reproduce the complete dynamics and morphology of the fluid in a cloud, but rather to assess the general characteristics of the whole MC ensemble.

The second caveat might be the assumption of steady state. Many authors find the latter as an idealization and this is certainly true for the most part of the MC life-cycle. Nevertheless, the steady-state assumption can be substantiated from consideration of time-scales. The changes of the large-scale velocity field (accretion) are slow; the internal sound crossing time near to the core and, hence, the time to establish local pressure equilibrium is much shorter. We suppose that there exists a period of time when the cloud reaches a nearly steady state \citep{Burkert_2017} and its PDF is characterized by two PLTs in the high density range, with roughly constant slopes \citep{KNW_2011,Girichidis_ea_2014,Schneider_ea_2015c}. Our model reproduces those two slopes in approximate solutions (Paper I, Paper II, this work). It should be added at the end that we consider the assumption of steady state as a proper first-step simplification in order to explain the limiting values of the slopes of power-law PDFs in late stages of MC evolution.

% Within the characteristic time of accretion the maximum increment of $M_0$ is of order of the total mass of the shells. 

Also, a question may arise whether the increase of the core mass $M_0$ due to accretion would not alter the energy balance, i.e. whether it would be in conflict with the assumption that the system is in (quasi-)steady state condition. This issue can be addressed as follows. Equation (\ref{Ealg_rho_G1}) may have solutions if the coefficients $A_0$, $3G_0$, $G_1$ and $2\Gamma/(\Gamma-1)$ are of the same order of magnitude. The first three are defined for the shells far from the core, at the outer edge of the cloud, while the coefficient $2\Gamma/(\Gamma-1)$ stems from the assumption of `hard polytropic' EOS near to the core. In this framework the physics of the outer shells (far from the core) determines not only their own dynamical state (like in Paper I and II) but also the state(s) of the shells near to the core. Therefore, to assess how the accretion flow affects the core mass $M_0$, one has to calculate the accretion rate through the outer cloud shells. Let $\dot{M}=dM/dt= 4\pi l^2 \rho u_{\rm a}$ be the accretion rate through the cloud shells at an arbitrary scale (radius) $l$. In view of the assumption for steady state, $\dot{M}$ is constant in the two main spatial regimes: far from and near to the core. In the former regime, we obtain a density profile $\rho(l)=\rho_{\rm c}(l/l_{\rm c})^{-2}$ and hence, from the continuity equation (formula \ref{va2-rho(l)}), a profile for the accretion velocity $u_{\rm a}=u_{\rm ac}={\rm const}$ with $u_{\rm ac}$ being the accretion velocity at the outer cloud edge. In the regime near to the core the two profiles are $\rho(l)=\rho_{\rm c}(l/l_{\rm c})^{-3}$ and $u_{\rm a}= u_{\rm ac} (l/l_{\rm c})$, respectively. After simple calculations, one obtains in both cases: $\dot{M}=4\pi l_{\rm c}^2 \rho_{\rm c} u_{\rm ac}=3M_{\rm c}^{*} (u_{\rm ac}/l_{\rm c})$ where $M_{\rm c}^{*}=(4\pi/3)l_{\rm c}^3 \rho_{\rm c}$ (introduced in Section \ref{Subsec-equ_rho(l)_terms} as a normalizing coefficient) is the mass of a cloud with size $l_{\rm c}$ and average density $\rho_{\rm c}$. It can be demonstrated that the averaged density of the outer shells is $\langle\rho\rangle_{\rm sh}\approx3\rho_{\rm c}$ (see Paper I, Section 4.1) and hence one gets for their total mass $M_{\rm sh}\approx 3M_{\rm c}^{*}$ and for the accretion rate $\dot{M}\approx M_{\rm sh} (u_{\rm ac}/l_{\rm c})= {\rm const}$. Now we define the characteristic accretion time scale as the time for which the total mass of the outer cloud shells flows down into the core: $\tau_{\rm a}=l_{\rm c}/u_{\rm ac}$; the total cloud size $l_{\rm c}$ is relevant here because the outer shells exceed significantly in size and volume the inner ones and the core.  Let us  compare $\tau_{\rm a}$ with the time scale $\tau_{\rm s}=l_{\rm c}/c_{\rm s}$ which is representative for the thermodynamic processes in the isothermal outer shells: $\tau_{\rm s}/\tau_{\rm a}=u_{\rm ac}/c_{\rm s}=\sqrt{A_0}\sim 1$ (see the definition of $A_0$ in Section \ref{Subsec-equ_rho(l)_terms}). The representative time scale for the gravity of the outer shells is the free-fall time: $\tau_{\rm ff}=\sqrt{3\pi/(32G\langle\rho\rangle_{\rm sh})}$. It can be easily shown that $\tau_{\rm ff}=1.57\tau_{\rm s}/\sqrt{3G_0}\sim\tau_{\rm s}$, since $\sqrt{3G_0}\approx2$. Hence the considered three characteristic time scales are of the same order of magnitude. Finally one obtains that the coefficient $G_1$ preserves its order of magnitude if the core mass increases (for one characteristic accretion time) with $M_{\rm sh}$ since the increment is $\sim3G_0$, and $3G_0 \sim A_0 \sim G_1$. We conclude that if there is a balance between the coefficients in equation (\ref{Ealg_rho_G1}) at the starting point of our consideration, this balance will be largely preserved after few accretion dynamical times and this is a necessary condition the (quasi-)steady state to hold. In other words, accretion, thermodynamics and gravity in the modeled cloud are of the same order of magnitude in terms of energies (per unit mass) and time scales.

A further justification of the (quasi-)steady state condition comes from comparison of the characteristic time for core mass' growth $\tau_0= M_0/\dot{M}$ with $\tau_{\rm a}$, $\tau_{\rm s}$, or $\tau_{\rm ff}$. Making use of the estimate $\dot{M}= 3M_{\rm c}^{*}/\tau_{\rm a}$ obtained above, one gets $\tau_0= (M_0/3M_{\rm c}^{*}) \tau_{\rm a}$. Per definition, we have $M_0= (4\pi/3) l_0^3 \rho_0$ and $M_{\rm c}^{*}=(4\pi/3) l_{\rm c}^3 \rho_{\rm c}$. Now, in view of the density profile $\rho(l)=\rho_{\rm c}(l/l_{\rm c})^{-2}$ in the shells from the cloud edge down to the close vicinity of the core (where the EOS changes), we obtain  $\rho_0 l_0^2 \approx \rho_{\rm c} l_{\rm c}^2$ and $M_0/3M_{\rm c}^{*} \approx (1/3) l_0/l_{\rm c} \ll 1$. Thus the characteristic time at which the core mass grows is $\tau_0 \ll \tau_{\rm a}$, and so the (quasi-)steady state assumption is fully justified. 

\section{Conclusions}
\label{Sec-conclusions}

In the presented paper we obtain an equation for the density profile function in a self-gravitating polytropic spherically symmetric turbulent fluid. It is appropriate to describe the conditions near to the core of an averaged representative of a whole class of molecular clouds (MCs) with one and the same PDF, sizes, boundary densities and temperatures \citep{DVK_2017}. The model is a modification of the framework established in \citet{DS_2018, DS_2019} changing the thermodynamics of the gas near to the very small and dense core from an isothermal ($\Gamma=1$) to a polytropic equation of state with index $\Gamma>1$. Assuming steady state in the cloud as the matter accretes through its outer boundary and passes through all spatial scales down to the core, we show that the total energy per unit mass is an invariant with respect to the fluid flow. The obtained non-linear integral equation for the density profile function reproduces the Bernoulli equation for our model and describes the balance of the kinetic, thermal and gravitational energy of a fluid element. 

We propose also a method to obtain an approximate solution for the density profile $\rho(l)\propto l^{-p}$. This allows us to derive a power-law PDF in the dense cloud regions (where self-gravity is important) as found from observations \citep{Schneider_ea_2015c} and numerical simulations \citep{KNW_2011, Veltchev_ea_2019}. Four solutions are obtained under different assumptions for the importance of the accretion kinetic term, thermal term or both with respect to the gravitational energy. The solutions correspond to different density profiles, polytropic exponents and energy balance equations for a fluid element. Two of the obtained solutions ($p=3/2$, $1<\Gamma<4/3$) and ($p=3$, $\Gamma=4/3$) can be interpreted as descriptions of contiguous stages of the evolution of a second power-law tail. The second solution applies to the detection of such power-law tail in PDFs in some observational and numerical studies of dense cloud regions \citep{Schneider_ea_2015c, KNW_2011, Marinkova_ea_2021}. The first power-law tail in PDFs at lower densities can be assessed from the consideration of the isothermal outer shells \citep{DS_2018, DS_2019}.

We are aware that this offers only a possible explanation of the second power-law tail in the PDF. Rotation of the core, magnetic fields and feedback from new-born stars may also play a significant role for this phenomenon. Therefore the emergence of a second power-law tail requires a more complex theoretical description and our model should be considered as a first simple step towards it. 

{\it Acknowledgement:}  
We are grateful to the anonymous referee for the valuable comments and suggestions which helped us to improve this paper. T.V. acknowledges support by the DFG under grant KL 1358/20-3 and additional funding from the Ministry of Education and Science of the Republic of Bulgaria, National RI Roadmap Project DO1-383/18.12.2020. R.S.K. acknowledges financial support from the German Research Foundation (DFG) via the Collaborative Research Center (SFB 881, Project-ID 138713538) 'The Milky Way System' (sub-projects A1, B1, B2, and B8). He also thanks for funding from the Heidelberg Cluster of Excellence STRUCTURES in the framework of Germany's Excellence Strategy (grant EXC-2181/1 - 390900948) and for funding from the European Research Council via the ERC Synergy Grant ECOGAL (grant 855130).

{\it Data availability:}
No new data were generated or analysed in support of this research.

\label{lastpage}

\end{document}